\documentclass[twocolumn,amsmath,amssymb,prb,draft]{revtex4}
\usepackage{latexsym}

\begin{document}

\author{Sandro Silva e Costa}
\title{The harmonic oscillator, dimensional analysis and inflationary solutions}
\date{December 5, 2002}

\affiliation{Instituto de Astronomia, Geof\'\i sica e Ci\^encias
Atmosf\'ericas\\ Universidade de
S\~ao Paulo\\ R. do Mat\~ao, 1226 -- Cidade Universit\'aria\\
CEP 05508-900 -- S\~ao Paulo -- SP -- Brazil}

\begin{abstract}
In this work, focused on the production of exact inflationary solutions
using dimensional analysis, it is shown how to explain inflation from a
pragmatic and basic point of view, in a step-by-step process, starting from
the one-dimensional harmonic oscillator.
\end{abstract}

\maketitle

\section{Introduction}

``No discussion of modern cosmology would be complete without some mention
of inflation .... This is the idea that ... there existed a phase in
which the universe expands much faster (in fact exponentially) than the rate
given by the standard scenario'' \cite{D'Inverno0}. Although now more than
20 years old, this idea survived well in the modern cosmology:
``cosmologists ... fell in love with inflation, because it could explain
some nagging problems raised by the standard big bang model'' \cite{Horgan}.
In fact, inflation is now probably more alive than ever, since the most
recent measurements of the cosmic microwave background indicate an almost
flat universe, a result {\it predicted} by inflation. But despite the
success of inflation among cosmologists, and several attempts to explain it,
inflation is only a general idea, since no one came up with a definitive
inflationary scenario.

As an important part of the modern cosmology, inflation is now a subject
present in several introductory texts, articles and books. In one science
dictionary, for example, one can read, under the entry ``early universe'',
that ``an important idea in the theory of the early universe is that of {\it %
inflation} -- the idea that the nature of the vacuum state gave rise, after
the big bang, to an exponential expansion of the universe'' \cite{Dictionary}%
. This shows that is common to associate the inflationary idea with a
`decay' or `phase transition' of the vacuum (or something alike), and it is
almost canonical to present inflation in this way \cite{Kolb}.

To be more specific, the most usual approach is to show inflation as an effect due to the
dynamics of a scalar field, with use of an approximation known as `{\it slow
rollover'}. However, from a practical point of view, it is possible to
present the construction of inflationary solutions without the mention or
use of any approximations, from very basic concepts. 
The dynamics of one single homogeneous scalar field is one-dimensional,
and as such it has similarities with the dynamics of the one-dimensional
harmonic oscillator, and this fact can be used in an introductory approach
to the idea of inflation, together with `suggestions' given by dimensional
analysis.

It is the purpose of this work, then, to show how to explain inflation from
a pragmatic and basic point of view, in a step-by-step process, starting
from the harmonic oscillator to finally get, after some general mathematical
efforts, {\it exact} inflationary solutions. Therefore, it must be noted
that here the focus is on the production of inflationary solutions, and not
on interpreting what they can mean or what kind of problems they can solve.

It is important to notice that, though pedagogical, this text takes
advantage of common conventions of the scientific notation. For example,
throughout this text natural units are used, ie, $c=G=\hbar =1$, unless
stated otherwise, like in the section of dimensional analysis. Also, the 
{\it sum convention} is always present, ie, terms like $\sum_{\mu
=0}^3\left( x^\mu \right) ^2$ are written as $x^\mu x_\mu $, with greek
indices ($\mu ,\nu ,\ldots $) running from $0$ to $3$ and latin ones ($%
i,j,\ldots $) from $1$ to $3$. Finally, generic constants are written
as $c_i$ ($ c_1, c_2, \ldots $) and $\lambda _i$ ($\lambda _1,
\lambda _2, \ldots $).

\section{The harmonic oscillator}

A basic {\it one-dimensional} system may be constructed with a particle of
mass $m$ and velocity $v$ subject to a force $F$, from which one obtains a 
{\it potential energy} $V\left( x\right) $, so that $V\left( x\right) =-\int
Fdx$. Such system may be described by the {\it Lagrangian} 
\begin{equation}
\label{Hm}L\left( x,v\right) =T-V\left( x\right) =\frac{mv^2}2-V\left(
x\right) ,
\end{equation}
where $T$ is the kinetic energy of the particle. Notice that the Lagrangian
is a function of the coordinate $x$ and its first derivative, the velocity $v
$, ie, $L=L\left( x,v\right) $.

From the Lagrangian one can build another function, the {\it Hamiltonian} 
\begin{equation}
\label{H}H\equiv pv-L=\frac{p^2}{2m}+V\left( x\right) {\cal ,}
\end{equation}
where 
\begin{equation}
\label{p}p\equiv \frac{\partial L}{\partial v}=mv
\end{equation}
is the momentum associated with the coordinate $x$. The Hamiltonian, which is
a function of $x$ and $p$, represents the total energy of the system.

If the total energy of the system is conserved, then 
\begin{equation}
\label{dHdt}\frac{dH}{dt}=\frac{\partial H}{\partial x}\left( \frac{dx}{dt}%
\right) +\frac{\partial H}{\partial p}\left( \frac{dp}{dt}\right) +\frac{%
\partial H}{\partial t}=0,
\end{equation}
or, equivalently,
\begin{equation}
\label{mv2}\frac d{dt}\left( \frac{mv^2}2+V\right) =0,
\end{equation}
from where one gets the equation of movement
\begin{equation}
\label{mdVdx}m\frac{d^2x}{dt^2}+\frac{dV}{dx}=0.
\end{equation}
What kind of potential can one have in this equation? {\it Possible} answers
can be given by dimensional analysis. Noting that 
\begin{equation}
\label{Vkgm}\left[ V\right] =kg^1m^2s^{-2}
\end{equation}
where $\left[ Q\right] $ means the dimension of the quantity $Q$, and $kg$, $%
m$ and $s$ are the units of mass, distance and time in SI units,
respectively, relations like
\begin{equation}
\label{Vlx}V=\lambda _1x^p
\end{equation}
are possible only if
\begin{equation}
\label{l1}\left[ \lambda _1\right] =\left[ Vx^{-p}\right] =kg^1m^{2-p}s^{-2}.
\end{equation}
Therefore, in this case the system must have some measurable constant $%
\lambda _1$ with dimensions given by (\ref{l1}). 

``The most important problem in one-dimensional motion ... is the harmonic
oscillator'' \cite{Symon}. This system may be represented by a mass fastened
to a spring. The spring is characterized by a constant $k$ which has
\begin{equation}
\label{k}\left[ k\right] =kg^1s^{-2},
\end{equation}
so that, for the harmonic oscillator, the correct form of the potential is
\begin{equation}
\label{vx}V\left( x\right) =\lambda _1x^2=\mu kx^2,
\end{equation}
where $\mu $ is a dimensionless constant. Since the spring obeys Hooke's law 
$F=-kx$, then $\mu =1/2$, and the equation of movement for the harmonic
oscillator is
\begin{equation}
\label{edm}m\frac{d^2x}{dt^2}+kx=0.
\end{equation}

The one-dimensional harmonic oscillator, which by its simplicity represents
a prototype for many one-dimensional systems, can be generalized in the
context of a field theory in a 4-dimensional -- or {\it covariant} --
formalism. Such generalization is the main topic of the next section,
where the Lagrangian formulation is also written covariantly.

\section{A generalized harmonic oscillator}

For a covariant generalization of the harmonic oscillator one needs some new
concepts, usually presented as a `preface to curvature' \cite{Schutz}:

\begin{itemize}
\item  the idea of a space-time, with coordinates 
$x^\mu $ and the metric $%
g_{\mu \nu }$ (of determinant $g$), described by the interval $%
ds^2\equiv g_{\mu \nu }dx^\mu dx^\nu $;

\item  the covariant derivative $\nabla _\mu $, which substitutes the usual
derivative $\partial _\mu \equiv d/{dx^\mu }$, and related to it by a
conexion $\Gamma _{\beta \gamma }^\alpha $, such that $\left( \nabla _\mu
-\partial _\mu \right) x^\nu =\Gamma _{\mu \alpha }^\nu x^\alpha $ {\it and} 
$\nabla _\mu g_{\alpha \beta }=0$;

\item  the idea of a\ {\it Lagrangian density} ${\cal L}={\cal L}\left(
x^\mu ,\nabla _\mu \varphi _i\right) $, which is a {\it scalar density} \cite
{D'Inverno} of weight $+1$, and which depends on the space-time coordinates $%
x^\mu $ through {\it fields} $\varphi _i\left( x^\mu \right) $ and their
gradients $\nabla _\mu \varphi _i\left( x^\mu \right) $.
\end{itemize}

Using these concepts, the simplest covariant field generalization of the
harmonic oscillator is 
\begin{equation}
\label{Lag}{\cal L}=\sqrt{-g}\left[ \frac 12g^{\mu \nu }\nabla _\mu \varphi
\nabla _\nu \varphi -V\left( \varphi \right) \right] . 
\end{equation}

Now, for {\it any} infinite system described by a general Lagrangian density 
${\cal L}$, one can write an {\it energy-momentum tensor }\cite{Birrel} 
\begin{equation}
\label{Tmunu}T^{\mu \nu }\equiv \frac 1{\sqrt{-g}}\left[ \pi ^\mu \nabla
^\nu \varphi _i-g^{\mu \nu }{\cal L}\right] , 
\end{equation}
where 
\begin{equation}
\label{pmu}\pi ^\mu \equiv \frac{\partial {\cal L}}{\partial \left( \nabla
_\mu \varphi \right) }, 
\end{equation}
and from it, by use of the conservation condition 
\begin{equation}
\label{cons}\nabla _\mu T^{\mu \nu }=0, 
\end{equation}
to obtain the equations of movement -- the {\it field equations} -- of the
system.

In the case of the generalized harmonic oscillator the field equation
obtained is the Klein-Gordon equation, 
\begin{equation}
\label{um}\Box \varphi +\frac{dV}{d\varphi }\equiv g^{\mu \nu }\nabla _\mu
\nabla _\nu \varphi +\frac{dV}{d\varphi }=0, 
\end{equation}
where $\Box $ is the d'Alembertian operator.
Notice that both the covariant derivative {\it and} the
d'Alembertian depend on the metric $g_{\mu \nu }$, and that one can write
for a scalar, explicitly \cite{Birrel2}, 
\begin{equation}
\label{Dal}\Box \varphi =\left[ -g\right] ^{-1/2}\partial _\mu \left( \sqrt{%
-g}g^{\mu \nu }\partial _\nu \varphi \right) . 
\end{equation}

Everything done in this section has a good analogy with the classical
description of the one-dimensional harmonic oscillator of the previous
section. For example, the one-dimensional harmonic oscillator has as
equation of movement a linear second-order differential equation, while the
field equation for the generalized harmonic oscillator is a generalization
of this, with the d'Alembertian operator acting as the second-order
differential operator. However, the d'Alembertian depends on the metric $%
g_{\mu \nu }$ of the spacetime. So, in order to solve the field equation for
the generalized harmonic oscillator first one must find how to write it
explicitly. This is done in the next section with the geometric formalism of
General Relativity.

\section{General Relativity}

The relation between the metric of a space-time and its content of matter 
{\it and} energy is given by the Einstein Equations 
\begin{equation}
\label{Gmunu}G_{\mu \nu }\equiv R_{\mu \nu }-\frac 12g_{\mu \nu }R=8\pi
T_{\mu \nu }+g_{\mu \nu }\Lambda , 
\end{equation}
where $R_{\mu \nu }$, the Ricci tensor, is a geometrical quantity built with
the metric and its derivatives up to second order, $R\equiv g^{\mu \nu
}R_{\mu \nu }$ is the Ricci scalar, and $\Lambda $ is the cosmological
constant. If one wants a very general solution of Einstein Equations,
representing an isotropic and homogeneous space-time, it is common to use
the metric present in the Friedmann--Lema\^\i tre-Robertson-Walker (FLRW)
element of line 
\begin{equation}
\label{FLRW}ds^2=dt^2-a^2\left( t\right) \left[ \frac{dr^2}{1-kr^2}%
+r^2\left( d\theta ^2+\sin ^2\theta d\phi ^2\right) \right] , 
\end{equation}
where $a\left( t\right) $ is the {\it scale factor}, and where $k=0,\pm 1$
is a number, the {\it parameter of curvature}. Another important assumption
is that the matter content of the universe may be represented by a generic
perfect fluid with 
\begin{equation}
\label{T2}T^{\mu \nu }=\left( \rho +p\right) u^\mu u^\nu -pg^{\mu \nu } 
\end{equation}
as energy-momentum tensor, where $u^\mu $ is the 4-velocity of the fluid, $%
\rho $ its energy density and $p$ its pressure. Using these ingredients, one
obtains two equations, one relating $\rho $ and $a$, 
\begin{equation}
\label{Friedmann}\frac 1{a^2}\left( \frac{da}{dt}\right) ^2+\frac k{a^2}=%
\frac{8\pi }3\rho +\frac \Lambda 3, 
\end{equation}
known as Friedmann's Equation, and another relating $p$ and $a$, 
\begin{equation}
\label{Ray}\frac 2a\frac{d^2a}{dt^2}+\frac 1{a^2}\left( \frac{da}{dt}\right)
^2+\frac k{a^2}=\Lambda -8\pi p. 
\end{equation}
It is important to notice that these two equations are not independent, due
to the conservation of energy, 
\begin{equation}
\label{dEpdV}d\left( \rho V\right) +pdV=0, 
\end{equation}
which relates the variation of $\rho $ with $p$ and the variation of the
volume $V$ of the fluid.

The important thing to be noticed here is that the metric $g_{\mu \nu }$ has
a time dependence, in the scale factor $a\left( t\right) $, represented by
any of the equations (\ref{Friedmann}) and (\ref{Ray}). Therefore, different
contents of matter and energy, expressed in $\rho $ and $p$, give birth to
different temporal evolutions of the metric. One kind of these possible
evolutions received the name of inflation.

\section{The field equations}

The FLRW metric also allows to write explicitly the d'Alembertian, through
eq. (\ref{Dal}), such that the field equation becomes

\begin{equation}
\label{field}\frac{\partial ^2\varphi }{\partial t^2}+\frac 3a\frac{da}{dt}%
\frac{\partial \varphi }{\partial t}+\frac 1{a^2}\nabla ^2\varphi +\frac{dV}{%
d\varphi }=0, 
\end{equation}
where $\nabla ^2$ is the usual three-dimensional Laplacian operator.

Can the energy-momentum tensor of the scalar field, obtained from equations (%
\ref{Lag}) and (\ref{Tmunu}), be compared to the one of a perfect fluid? If
one assumes that the field is homogeneous there is no spatial gradients and,
therefore, from the component $T^{00}$, one gets 
\begin{equation}
\label{T00}\rho =\frac 12\left( \frac{d\varphi }{dt}\right) ^2+V\left(
\varphi \right) 
\end{equation}
and, after this, it becomes clear that 
\begin{equation}
\label{Tii}p=\frac 12\left( \frac{d\varphi }{dt}\right) ^2-V\left( \varphi
\right) . 
\end{equation}
This gives the {\it equation of state} 
\begin{equation}
\label{EDE}p=w\rho , 
\end{equation}
where 
\begin{equation}
\label{w}w\equiv \left[ \frac 12\left( \frac{d\varphi }{dt}\right)
^2-V\left( \varphi \right) \right] \left[ \frac 12\left( \frac{d\varphi }{dt}%
\right) ^2+V\left( \varphi \right) \right] ^{-1}. 
\end{equation}

Using the above results one obtains two independent equations, 
\begin{equation}
\label{dois}\left( \frac 1a\frac{da}{dt}\right) ^2+\frac k{a^2}=\frac{8\pi }%
3\left[ \frac 12\left( \frac{d\varphi }{dt}\right) ^2+V\left( \varphi
\right) \right] +\frac \Lambda 3, 
\end{equation}
or its equivalent 
\begin{equation}
\label{doisa}\frac 2a\frac{d^2a}{dt^2}+\frac 1{a^2}\left( \frac{da}{dt}%
\right) ^2+\frac k{a^2}=\Lambda -8\pi \left[ \frac 12\left( \frac{d\varphi }{%
dt}\right) ^2-V\left( \varphi \right) \right] , 
\end{equation}
and 
\begin{equation}
\label{tres}\frac{d^2\varphi }{dt^2}+\frac 3a\frac{da}{dt}\frac{d\varphi }{dt%
}+\frac{dV}{d\varphi }=0, 
\end{equation}
which becomes the equation for the harmonic oscillator if $a$ is constant,
or its equivalent 
\begin{equation}
\label{tresa}\frac d{dt}\left[ \frac 12\left( \frac{d\varphi }{dt}\right)
^2+V\left( \varphi \right) \right] +\frac 3a\frac{da}{dt}\left( \frac{%
d\varphi }{dt}\right) ^2=0, 
\end{equation}
which is nothing more than the condition for energy conservation, with $a^3$
assuming the role of the volume.

Unfortunately, the {\it two} independent equations obtained contain {\it three}
unknowns, $a$, $\varphi $ and $V$. Therefore, in order to solve them one
needs something else, like simplifying assumptions, or an equation
relating $V$ and $\varphi $, for example. The most common procedure is to
specify some kind of potential, $V\left( \varphi \right) $, and from it
solve the equations for the field and the scale factor. In the next section
it is shown that a group of inflationary solutions can be built with
relations suggested by dimensional analysis.

\section{Dimensional considerations}

A natural choice to solve the system formed by the equations (\ref{dois})
and (\ref{tres}) consists in using as an extra hypothesis some relation
between the unknowns $a$ and $\varphi $ (and their derivatives) suggested by
dimensional analysis. As a specific example, one can use only relations
where all needed constants are built with a combination of the universal
constants $c$ and $G$, both present in General Relativity.

The first step consists in observing that 
\begin{equation}
\label{Gcs}\left[ G^pc^q\right] =kg^{-p}m^{3p+q}s^{-2p-q}.
\end{equation}
Assuming 
\begin{equation}
\label{am}\left[ a\right] =m
\end{equation}
and 
\begin{equation}
\label{fiponto}\left[ \left( \frac{d\varphi }{dt}\right) ^2\right]
=kg^1m^{-1}s^{-2},
\end{equation}
one can `try' relations such as 
\begin{equation}
\label{quatroa}\left( \frac{d\varphi }{dt}\right) ^{2k}=\lambda _1a^{-\ell },
\end{equation}
where 
\begin{equation}
\label{quatro}\left[ \lambda _1\right] =\left[ \left( \frac{d\varphi }{dt}%
\right) ^{2k}a^\ell \right] =kg^km^{-k+\ell }s^{-2k}.
\end{equation}
Comparing this last result with equation (\ref{Gcs}) one obtains the system
of equations 
\begin{equation}
\label{kpq}\left\{ 
\begin{array}{l}
k=-p \\ 
-k+\ell =3p+q \\ 
-2k=-2p-q
\end{array}
\right. ,
\end{equation}
which gives $p=-k$, $q=4k$ and $\ell =2k$. Choosing $k=1/2$ one gets $\left[
\lambda _1\right] =\left[ G^{-1/2}c^2\right] $ and the relation 
\begin{equation}
\label{fiponto1}\frac{d\varphi }{dt}=\lambda _1a^{-1}.
\end{equation}
Substituting this `ansatz' in (\ref{dois}) and in (\ref{tresa}), one obtains
a new system formed by the equations 
\begin{equation}
\label{f1}\left( \frac 1a\frac{da}{dt}\right) ^2+\frac k{a^2}=\frac{8\pi }%
3\left[ \frac{\lambda _1^2}{2a^2}+V\left( \varphi \right) \right] +\frac
\Lambda 3
\end{equation}
and 
\begin{equation}
\label{dadV}\frac{2\lambda _1^2}{a^3}\frac{da}{dt}+\frac{dV}{dt}=0.
\end{equation}
This last equation shows that 
\begin{equation}
\label{Vv0}V=V_0+\lambda _1^2a^{-2},
\end{equation}
with $V_0$ being a constant, what finally allows one the obtain the
solutions 
\begin{equation}
\label{a1}a\left( t\right) =\frac 1{2\alpha }\left[ c_1e^{\alpha t}+\frac{%
k-4\pi \lambda _1^2}{c_1}e^{-\alpha t}\right] ,
\end{equation}
\begin{equation}
\label{fi1}\varphi \left( t\right) =\varphi _0+\frac{\lambda _1}{\left( 4\pi
\lambda _1^2-k\right) ^{1/2}}\ln \left[ \frac{\left( 4\pi \lambda
_1^2-k\right) ^{1/2}-c_1e^{\alpha t}}{\left( 4\pi \lambda _1^2-k\right)
^{1/2}+c_1e^{\alpha t}}\right] ,
\end{equation}
and 
\begin{equation}
\label{V1}V\left( \varphi \right) =V_0-\frac{\alpha ^2}{k-4\pi \lambda _1^2}%
\sinh ^2\left[ \frac{\left( 4\pi \lambda _1^2-k\right) ^{1/2}}{\lambda _1}%
\left( \varphi -\varphi _0\right) \right] ,
\end{equation}
where $c_1$, $\varphi _0$ and $3\alpha ^2\equiv 8\pi V_0+\Lambda $ are
constants. Notice that these solutions, which do not always include the
initial condition $a\left( 0\right) =0$, are {\it not} valid for the case $%
\alpha =0$, when one has instead 
\begin{equation}
\label{a1a}a\left( t\right) =\left( 4\pi \lambda _1^2-k\right) ^{1/2}t,
\end{equation}
\begin{equation}
\label{fi1a}\varphi \left( t\right) =\varphi _0+\lambda _1\left( 4\pi
\lambda _1^2-k\right) ^{-1/2}\ln t,
\end{equation}
and 
\begin{equation}
\label{V1a}V\left( \varphi \right) =-\frac \Lambda {8\pi }-\frac{\lambda _1^2%
}{k-4\pi \lambda _1^2}\exp \left[ -2\sqrt{4\pi -\frac k{\lambda _1^2}}\left(
\varphi -\varphi _0\right) \right] .
\end{equation}

Another possible relation inspired by dimensional analysis is 
\begin{equation}
\label{seisa}\left( \frac{d\varphi }{dt}\right) ^{2k}=\lambda _2H^n\equiv
\lambda _2\left( \frac 1a\frac{da}{dt}\right) ^n, 
\end{equation}
with $H$ being now the {\it Hubble function}, and 
\begin{equation}
\label{seis}\left[ \lambda _2\right] =\left[ \left( \frac{d\varphi }{dt}%
\right) ^{2k}H^{-n}\right] =kg^km^{-k}s^{-2k+n}. 
\end{equation}
Using (\ref{Gcs}) one has, then, $n=q=-2p=2k$. Choosing $k=1/2$, $\left[
\lambda _2\right] =\left[ G^{-1/2}c^1\right] $ and the {\it ansatz} now
becomes 
\begin{equation}
\label{dlna}\frac{d\varphi }{dt}=\lambda _2\frac 1a\frac{da}{dt}=\lambda _2%
\frac{d\ln a}{dt}, 
\end{equation}
or 
\begin{equation}
\label{aexp}a=\exp \left( \frac{\varphi -\varphi _0}{\lambda _2}\right) , 
\end{equation}
what substituted in (\ref{dois}) and in (\ref{tresa}) yields, for $k=\Lambda
=0$, 
\begin{equation}
\label{a2a}a\left( t\right) =t^{1/\left( 4\pi \lambda _2^2\right) }, 
\end{equation}
\begin{equation}
\label{fi2a}\varphi \left( t\right) =\varphi _0+\frac 1{4\pi \lambda _2}\ln
t, 
\end{equation}
and 
\begin{equation}
\label{V2}V\left( \varphi \right) =\left( \frac 3{8\pi \lambda _2^2}-\frac
12\right) \frac{\exp \left[ -8\pi \lambda _2\left( \varphi -\varphi
_0\right) \right] }{16\pi ^2\lambda _2^2},
\end{equation}
where $\varphi _0$ is a constant.

A last example is given by the relation 
\begin{equation}
\label{cincoa}\left( \frac{d\varphi }{dt}\right) ^{2k}=\lambda _3\left( 
\frac{dH}{dt}\right) ^j, 
\end{equation}
where 
\begin{equation}
\label{cinco}\left[ \lambda _3\right] =\left[ \left( \frac{d\varphi }{dt}%
\right) ^{2k}\left( \frac{dH}{dt}\right) ^{-j}\right] =kg^km^{-k}s^{-2k+2j}. 
\end{equation}
Now, the indices are such that $2j=q=-2p=2k$, and choosing $k=1$ one gets $%
\left[ \lambda _3\right] =\left[ G^{-1}c^2\right] $ and the relation 
\begin{equation}
\label{dHubbledt}\left( \frac{d\varphi }{dt}\right) ^2=\lambda _3\frac{dH}{dt%
}, 
\end{equation}
with the solutions 
\begin{equation}
\label{a3}a\left( t\right) =\frac 1{2\beta }\left[ c_2e^{\beta t}+\frac
1{c_2}\left( \frac k{1+4\pi \lambda _3}\right) e^{-\beta t}\right] , 
\end{equation}
\begin{equation}
\label{fi3}\varphi \left( t\right) =\varphi _0+\sqrt{-\lambda _3}\ln \left[ 
\frac{k^{1/2}\left( 1+4\pi \lambda _3\right) ^{-1/2}-c_2e^{\alpha t}}{%
k^{1/2}\left( 1+4\pi \lambda _3\right) ^{-1/2}+c_2e^{\alpha t}}\right] , 
\end{equation}
and 
\begin{equation}
\label{V3}V\left( \varphi \right) =V_0-\lambda _3\beta ^2\left( \frac
12+\cosh ^2\frac{\varphi -\varphi _0}{\sqrt{-\lambda _3}}\right) , 
\end{equation}
where $c_2$, $V_0$ and $\varphi _0$ are constants, and $3\beta ^2\equiv
\left( 8\pi V_0+\Lambda \right) \left( 1+4\pi \lambda _3\right) ^{-1}$.
These solutions, which do not always include the initial condition $a\left(
0\right) =0$, are valid for $\beta \neq 0$. When $\beta =0$ one has instead 
\begin{equation}
\label{a3a}a\left( t\right) =k^{1/2}\left( 1+4\pi \lambda _3\right)
^{-1/2}t, 
\end{equation}
\begin{equation}
\label{fi3a}\varphi \left( t\right) =\varphi _0+\sqrt{-\lambda _3}\ln t, 
\end{equation}
and 
\begin{equation}
\label{V3a}V\left( \varphi \right) =-\frac \Lambda {8\pi }-\lambda _3\exp
\left[ -2\frac{\left( \varphi -\varphi _0\right) }{\sqrt{-\lambda _3}}%
\right] . 
\end{equation}

It is interesting to notice that a relation like 
\begin{equation}
\label{sete}\left( \frac{d\varphi }{dt}\right) ^{2k}=\lambda _4\left( \frac{%
da}{dt}\right) ^q 
\end{equation}
is not valid if one wants $\lambda _4$ a constant with dimension equal to
the dimension of a combination of only $c$ and $G$.

To finish this section it is important to remark that the solutions obtained
above, all of them generalizations of solutions presented before in the
literature \cite{Ellis}, are
only of two kinds with relation to the behaviour of the scale factor $a$: a
power function of $t$ and a combination of exponentials of $t$. In the next
section it will be discussed if such kinds of solutions are inflationary
solutions.

\section{Inflation}

The inflationary epoch of the universe is usually seen as the age where the
universe inflated very fast, in an exponential way. However, this
`definition' of inflation can be widened: one can say that ``inflation is
defined as an epoch in the history of the universe during which the cosmic
expansion is accelerated'' \cite{Faraoni}. This means that a solution of
Friedmann's equation for the scale factor $a\left( t\right) $ can be seen as
an inflationary solution if it is an expanding solution {\it and} its second
time derivative is positive, 
\begin{equation}
\label{condition}\frac{d^2a}{dt^2}>0. 
\end{equation}
However, combining (\ref{Friedmann}) and (\ref{Ray}), one can see that 
\begin{equation}
\label{d2adt2}\frac{d^2a}{dt^2}=\frac a3\left[ \Lambda -4\pi \left( \rho
+3p\right) \right] , 
\end{equation}
so that, considering only solutions and times where $a\geq 0$, the condition
for inflation becomes equivalent to have 
\begin{equation}
\label{condition2}\Lambda >4\pi \left( \rho +3p\right) . 
\end{equation}
Defining a new density $\rho _\Lambda $ such that $8\pi \rho _\Lambda \equiv
\Lambda$ and assuming the equation of state $p=w\rho $ this relation is
written as 
\begin{equation}
\label{rhol}\frac{\rho _\Lambda }{\rho } >\frac{1+3w}{2}. 
\end{equation}
For ordinary matter (`dust') $w=0$, while for radiation $w=1/3$. If,
however, $\Lambda=0 $ then one must necessarily have $w<-1/3$.

The solutions presented in the previous section are of two kinds: a power
function of $t$, where 
\begin{equation}
\label{at1}a\left( t\right) =c_1t^\lambda \Rightarrow \frac{d^2a}{dt^2}=%
\frac{\lambda \left( \lambda -1\right) }{t^2}a\left( t\right) , 
\end{equation}
and a combination of exponentials of $t$, with 
\begin{equation}
\label{at2}a\left( t\right) =c_2e^{\lambda t}+c_3e^{-\lambda t}\Rightarrow 
\frac{d^2a}{dt^2}=\lambda ^2a\left( t\right) . 
\end{equation}
Clearly, such solutions obey the condition for inflation with two
restrictions: $\lambda >1$, in the power function case, and $\lambda ^2>0$,
in the exponential case. Therefore, from all solutions presented here, in
this section, only the ones given by equations (\ref{a1a}) and (\ref{a3a})
can not be considered inflationary, while the solution given by eq. (\ref{a1}%
) is inflationary for $\Lambda >-8\pi V_0$, the one given by eq. (\ref{a2a})
is inflationary for $1>4\pi \lambda _2^2$, and the one given by (\ref{a3})
is inflationary wherever $\left( 8\pi V_0+\Lambda \right) \left( 1+4\pi
\lambda _3\right) ^{-1}>0$.

\section{Final remarks}

The purpose of this work was to show, using as example the `idea' of
inflation, that even very basic concepts, such as the use of dimensional
analysis, can have its role in the search of more sophisticated results.
This process can be illustrated a little more with one last example. If one
tries the relation 
\begin{equation}
\label{farao1}\left( \frac{d\varphi }{dt}\right) ^{2k}=\lambda _2H^n+\lambda
_2^{\prime }, 
\end{equation}
one gets that 
\begin{equation}
\label{farao2}\left[ \lambda _2^{\prime }\right] =kg^km^{-k}s^{-2k}, 
\end{equation}
and certainly this can not be obtained using only $c$ and $G$. Now if one
allows the use of Planck's constant $\hbar $, one has 
\begin{equation}
\label{Gch}\left[ G^pc^q\hbar ^r\right] =kg^{-p+r}m^{3p+q+2r}s^{-2p-q-r}, 
\end{equation}
such that $2r=p=-2q/7=-2k$. Choosing $k=1$, one obtains $\left[ \lambda
_2^{\prime }\right] =\left[ G^{-2}c^8\left( \hbar c\right) ^{-1}\right] $
and 
\begin{equation}
\label{farao3}\left( \frac{d\varphi }{dt}\right) ^2=\lambda _2H^2+\lambda
_2^{\prime } 
\end{equation}
Deriving this relation and substituting the result in (\ref{tresa}) one gets 
\begin{equation}
\label{seila}\lambda _2H\frac{dH}{dt}+3H\left( \lambda _2H^2+\lambda
_2^{\prime }\right) +\frac{dV}{dt}=0. 
\end{equation}
Now, inserting (\ref{farao3}) in (\ref{dois})  
and returning the result into (\ref{seila}) one gets 
\begin{equation}
\label{farao8}\frac{dH}{dt}+4\pi \left( \lambda _2H^2+\lambda _2^{\prime
}\right) -\frac k{a^2}=0. 
\end{equation}
This equation is easily solved when $k=0$, with the results 
\begin{equation}
\label{farao9}H\left( t\right) =-\sqrt{\frac{\lambda _2^{\prime }}{\lambda _2%
}}\tan \left[ 4\pi \sqrt{\lambda _2\lambda _2^{\prime }}\left( t-t_0\right)
\right] 
\end{equation}
and 
\begin{equation}
\label{farao10}a\left( t\right) =a_0\cos ^{1/\left( 4\pi \lambda _2\right)
}\left[ 4\pi \sqrt{\lambda _2\lambda _2^{\prime }}\left( t-t_0\right)
\right] , 
\end{equation}
where $t_0$ and $a_0$ are constants of integration. Imposing that $a\left(
0\right) =0$ one gets $2\gamma t_0=\pi $, and if $\lambda _2>0$ and $\lambda
_2^{\prime }<0$, then $\sqrt{\lambda _2\lambda _2^{\prime }}$ is a purely
imaginary quantity, such that 
\begin{equation}
\label{farao12}a\left( t\right) =a_1\sinh ^{1/\left( 4\pi \lambda _2\right)
}4\pi \sqrt{\lambda _2\left| \lambda _2^{\prime }\right| }t, 
\end{equation}
where $a_1$ is a real positive constant. Finally, if one chooses $4\pi
\lambda _2=3$, then, from (\ref{dois}) and (\ref{farao3}), $V=V_0=\left( 4\pi \left| \lambda
_2^{\prime }\right| -\Lambda \right) (8\pi )^{-1}$ and 
\begin{equation}
\label{farao14}a\left( t\right) =a_1\sinh ^{1/3}\sqrt{3\left( 8\pi
V_0+\Lambda \right) }t. 
\end{equation}
Notice that this {\it exact} solution, obtained previously in the literature
through the supposition of a specific form for the potential $V$ of the
scalar field \cite{Faraoni}, is just a particular case of a more general
solution given by eq. (\ref{farao12}).

It is important to emphasize that if one uses dimensional analysis to obtain
new reasonable relations between physical quantities, relations which can be
used as working hypothesis, one must worry about the validity -- or origin
-- of such relations: do they represent something more than a possible
mathematical relation with the dimensions adjusted correctly? For example,
in the case worked above does the use of Planck's constant $\hbar $ to
obtain the right dimensions of the constant $\lambda _2^{\prime }$ indicate
that such constant must have a quantum origin?

However, putting aside interpretative questions, one message this work
wants to leave is that dimensional
analysis is a very helpful and handy tool, which beyond allowing to check the
consistency of some results, can also suggest new approaches, explanations
and results, both in teaching and research.

\section*{Acknowledgements}

S. Silva e Costa thanks the Brazilian agency FAPESP (Funda\c c\~ao de Amparo
\`a Pesquisa do Estado de S\~ao Paulo) for financial support (grant
00/13762-6).


\begin{thebibliography}{99}
\bibitem{D'Inverno0}  R. d'Inverno, {\it Introducing Einstein's Relativity}
(Clarendon, Oxford, 1995), p. 354.

\bibitem{Horgan}  J. Horgan, {\it The End of Science} (Abacus, London,
1998), p. 99.

\bibitem{Dictionary}  {\it Concise Science Dictionary} (Oxford, Oxford,
1991), pp. 214-215.

\bibitem{Kolb}  E.W. Kolb and M.S. Turner, {\it The Early Universe} (Addison
Wesley, Reading, MA, 1993), pp. 261-367.

\bibitem{Symon}  K.R. Symon, {\it Mechanics} (Addison-Wesley, Reading, MA,
1973), pp. 39.

\bibitem{Schutz}  B.F. Schutz, {\it A first course in general relativity}
(Cambridge, Cambridge, 1985), pp. 118-194.

\bibitem{D'Inverno}  R. d'Inverno, {\it op. cit.}, pp. 91-99.

\bibitem{Birrel}  N.D. Birrel and P.C.W. Davies, {\it Quantum fields in
curved space} (Cambridge, Cambridge, 1994), p. 87.

\bibitem{Birrel2}  N.D. Birrel and P.C.W. Davies, {\it op. cit.}, p. 38.

\bibitem{Ellis}  G.F.R. Ellis and M.S. Madsen, ``Exact scalar fields
cosmologies'', Class. Quant. Grav. {\bf 8}, 667-676 (1991).

\bibitem{Faraoni}  V. Faraoni, ``A new solution for inflation'', Am. J.
Phys. {\bf 69} (3), 372-376 (2001).
\end{thebibliography}
\end{document}